# Ultrasound and Temperature Study of Non-Equilibrium Phase Transitions in Surface-Bound Liquid Layers

V.A. Shulgin

*Voronezh State University*

**Abstract**

The present research deals with the registration and study of temperature and ultrasound parameters of non-equilibrium phase transitions occurring in the layer "liquid – solid surface" in the process of slow heating and cooling of the medium. Different resistive sensors have been used for taking measurements. As a result there has been discovered a periodic stepped dependence of the registered temperature with jumps ~7.5-13 K in water, water solutions and other liquids. We have studied the conditions under which the stepped temperature dependence of the resistive sensors indications is synchronous to the resonance frequency variations of the liquid volume limited by the surface of the axial piezoelectric resonator. The experiments showed that for studying temperature transitions and ultrasound phase velocity variations the following conditions are required: ac field in the area of the liquid surface layer. Assumed cause of the temperature jumps occurrence is high heat conduction and heat capacity of the surface-bound layer, causing local temperature variations on the surface of the resistive sensor in the process of evolution of the two-dimensional crystal liquid structure. Investigation of non-equilibrium transitions of the surface-bound liquid layer may be of interest for studying molecular complexes of liquid media, as well as cellular structures of living systems.

**Preface**

Many researchers concentrated their efforts on studying the processes occurring at the interface boundary. One of the most interesting subjects of these studies is water, which properties in boundary layers differ considerably from its bulk phase characteristics. This information may be important for solving a number of important practical problems. Nanostructure technology, surface effects study in biological environment are the most important directions requiring data on non-equilibrium dissipative processes occurring at the interface "liquid – solid surface".

Structural effects of bound water anomalous properties can be easily traced in temperature dependence experiments. Some effects of surface-bound water are generally known. These are rise of the near surface relative viscosity at low temperatures, freezing point depression of bound water as compared with free water, dissolving capacity depression in thin films, permittivity different from free water [1.2]. Bound water structural changes stipulate its thermal diffusivity change. Deviation of bound water thermal diffusivity in the line of reduction – as compared to free water – starts to appear in water films less than 1 μm thick. The thinner is the bound water layer – the lower is its thermal diffusivity. In interlayers 0.03 μm thick thermal diffusivity is approximately 30% less as compared to free water [3].

As we can conclude from experimental data, structural changes in liquids occur due to the temperature change. This postulate has been made the basis for selecting methods of analysis and tools for the execution of the present work. Special attention has been paid to arrangement of specific conditions for measuring the macroscopic properties (ultrasound temperature and velocity) in layers of water and other liquid media, bound by solid surface.

The test object was considered as an "open system" with dissipative loss balanced by energy inflow. Control parameters have been selected subject to the principles resulting in constructive evolution of the open system, i.e. one parameter (temperature) is wandering, the other parameters are constant, and the system is a macroscopic space-time structure, studied independently subject to the initial conditions. Ultrasound study of the liquid was implemented by the instrumentality of a multiple-beam interferometer [4] which recorded the driving frequency of standing waves in a liquid volume, filling an axial resonator.

This technique made it possible to record small changes of ultrasound phase velocity ($\sim 10^{-6}$). Analysis of the ultrasound studies showed the existence of recycling anomalous variations of phase velocity in the temperature range which caused more detailed experimental investigation presented in this paper.

Presumably, the effects under study are stipulated by the dissipative processes in the liquid layer on the surface of sensors – axial acoustical resonator and resistive temperature detector. To reveal the general laws, integrated temperature and ultrasound tests of surface-bound structural formations of liquid medium were carried out synchronously.

**Description of tools for temperature and ultrasound tests of a surface-bound liquid layer**

Test liquid at a certain temperature or ice in amount of 100 g was placed in a glass or plastic bowl. A plastic bowl of minimum weight (2.5 g) was used for carrying out calorimetric measurements. A resistive heater (5 O) up to 5 W power was placed into a minimum size plastic bag, in order to isolate it from the total volume. The plastic bag was filled with a small quantity of distilled water to provide thermal contact with the liquid.

The heater was placed into the liquid volume close to the bowl wall so that to form a circulating flow and corresponding mixing of the thermal medium. The DC power supply provided the calculated value of the heater power (1 – 4.5 W). For most of the temperature measurements a platinum resistive sensor of Honeywell production (HEL-775-B) was used. The sensor is a film structure ~ 1mm$^2$ in area on a ceramic substrate with protective dielectric coating. Sensor resistance is 100 O at $0°$C, absolute error – 0.6% of the temperature measured in the range of -55$°$ up to +150° C. Time constant of the sensor dipped into liquid was ~ 1s. The sensor was placed into the tested liquid at some distance from the heater. The measurements were carried out by RLS meter E7-22 (LCR 4080) with the measuring signal of alternating current (frequency 120 Hz, 1000 Hz; effective value of the measuring voltage 100 mV). Relative error of the measurement was ~ 0.5%. Resolution of the meter made it possible to register temperature changes with discreteness of 0.03 K. Data transmission was carried out via interface RS-232 with optron insulation. The test liquid was placed into a heat-insulating container supplied with a thermal radiation reflector made of aluminum foil.

Measurement time and temperature data were registered by a computer with an interval of 1 – 2 s and processed in Microsoft Office Excel. Ultrasound tests were carried out with the assistance of an oscillator for liquid media analysis [4], which resonance system can be defined as a multiple-beam acoustic interferometer. Radiative and receiving electroacoustic transducers of the interferometer form a component of the axial piezoelectric resonator (Fig. 1). The prototype of such open type resonator is the Fabry-Pérot optical interferometer. It is known that numerical values of the "transparency" frequencies of Fabry-Pérot interferometer depend specifically on the refractive index of the medium between input and output reflection planes, where the standing-wave mode is excited. Similarly, frequency eigenvalues of the liquid in the cylindrical interior of the resonator (Fig. 1) depend on the ultrasound phase velocity in the tested medium. One of these frequencies is separated by the filter and specifies electric oscillation. Feedback of the oscillator is closed up by the ultrasonic vibration propagation medium. Multiple-beam interference occurs as a result of the standing-wave mode excitement in the axial resonator. Oscillator frequency dispersion is restricted by the frequency range in the neighbourhood of piezoelectric resonator wall natural oscillations and the mode filter adjusted to the same frequency.

Similar to the Fabry-Pérot interferometer, in the process of many beams interference there occurs spatial redistribution of the energy flux with its concentration in one direction (radius of the axial resonator). The effect of coherent energy integration under the condition of phase synchronism is high Q factor of resonance in the liquid volume and, respectively, high stability of the oscillation frequency ($\sim 10^{-6}$). Signal level increase on the resonance frequency is rated by the AGC system which results in frequency suppression and disagreement with the condition of phase synchronism. AGC threshold is selected basing on the condition of the oscillation low-level signal operation. In the present paper the exciting voltage of the structure was 0.3 V at the frequency ~1.7 MHz. Change of the medium state, i.e. change of the phase delay of longitudinal ultrasonic waves propagation, results in standing waves eigenfrequency change. In the range of frequency dispersion there occur oscillation frequency variations whereas the above range is restricted to the frequency band of the mode filter and natural resonance of the piezoelectric resonator wall thickness.

Geometry features of the resonator are: height – 13 mm, inside diameter – 16 mm, wall thickness – 1.25 mm. In case of big variations, oscillated frequency may be beyond the scope of free spectral range in the current order of interference and transfer to the adjacent order. There may occur several transitions of this kind.

Axial construction of the resonator is suitable for thermostating. Besides, the ultrasound field of standing waves is concentrated inside the cylinder, and the walls of the bowl with liquid under study exert minor influence on the oscillator frequency variations. It should be noted that due to its high frequency resolution (i.e. high sensitivity to variations of the medium physical characteristics) acoustic interferometer can register changes in the liquid thin layer against integral characteristic of the medium.

Data on the frequency of electric oscillations was transferred to the computer with 1s interval and registered jointly with the temperature data. Registration nonsynchronism amounted to 1 – 2s. Data was processed in Microsoft Office Excel.

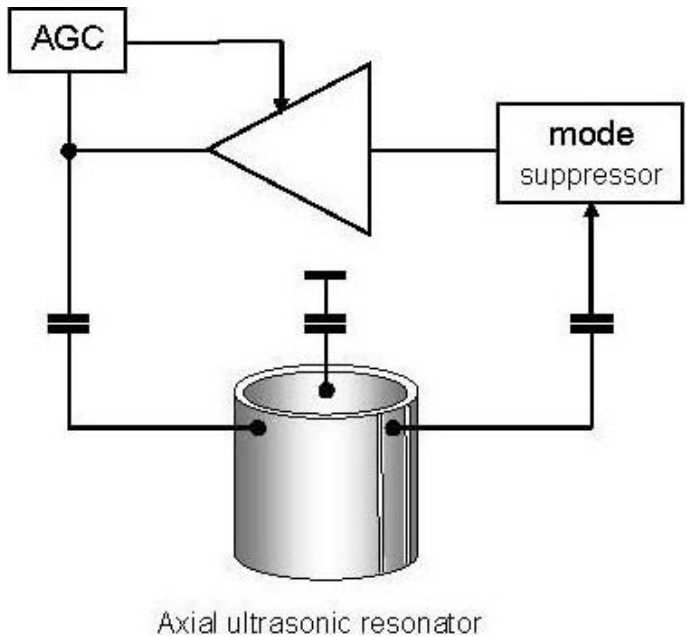

Fig. 1 Oscillator of ultrasonic vibration in liquid

**Data on temperature and ultrasonic tests of surface-bound layer of liquid media**

In the process of slow heating and cooling a water volume we have discovered stepped changes in resistive sensor temperature indications. Fig. 2 presents a temperature-time diagram (100 ml of distilled water was heated in a thermally insulated container by a 3 W power heater). Simultaneously we made measurements with a K-type thermocouple (diameter 0.25 mm) of a digital meter VC9808. Film sensor (measuring signal is ac voltage) registered slow temperature rise (~ 0.003 K/s) with repeating rapid rises of 7.5 – 8.5 K at a speed of ~ 2 – 8 K/s. Thermocouple sensor registered uniform temperature dependence in the form of an envelope of the stepped signal.

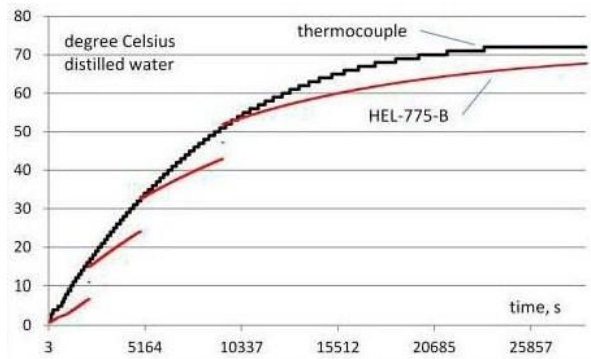

Fig. 2 Temperature – time diagram (100 ml of distilled water heated by a 3W power heater, measurements made by a K-type thermocouple (0.25 mm) and a resistive sensor HEL-775-B)

An attempt was made to disclose anomalous temperature dependence, mentioned in the paper [5] in similar measurement conditions. In the process of experiment (Fig.3), in which an L-type thermocouple (0.1 mm) was used, there have been observed deviations from uniform dependence in case of slow heating. The diagram illustrates the existence of random component in temperature indications. Difference in the dynamics of the heating process development in case of similar media and similar initial conditions (Fig. 2, 3) is stipulated by different heat loss due to different diameter of thermocouples. When temperature was registered by sensor HEL-775-B (measuring signal was dc voltage, meter VC9808), usual continuous temperature dependence of uniformly heated medium was observed.

Results of this research indicate that, first of all, it is necessary to take into consideration physical conditions of sensors performance (geometrics, fields' availability, medium properties on the sensors surface).

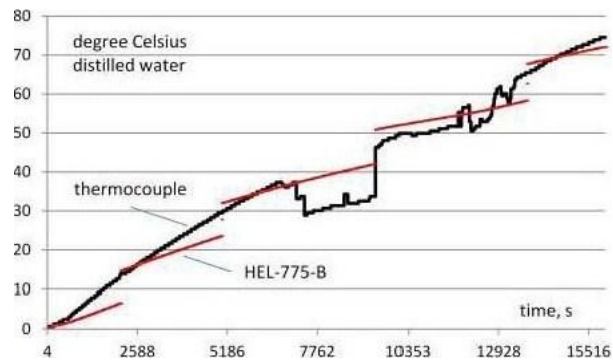

Fig. 3 Temperature – time diagram (100 ml of distilled water heated by a 3W power heater, measurements made by an L-type thermocouple (0.1 mm) and a film (Pt) resistive sensor, measuring signal – ac voltage)

A thermocouple generates a local electric field in the area of junction. Meter E7-22 forms variable electric field in the surface-bound water layer; this is realized by use of a film sensor HEL-775-B. Meter VC9808 measures resistance at low current and low voltage drop on the resistive structure. Experiments demonstrated that stepped dependence of temperature indications can be always observed when there is ac voltage on the resistive sensor, which is galvanically isolated from external facilities.

Alongside with periodic variations of temperature indications, in the process of carrying out an experiment in accordance with the above technique there have been detected synchronous variations of ultrasonic parameters of liquid. Fig. 4 presents a diagram of synchronous measurements of temperature (100 ml of distilled water heated by a 3W power heater) and eigenfrequency of liquid resonance in axial resonator.

Break points of oscillation frequency-temperature dependence correspond to the transitions into adjacent orders of interference. The diagram would be continuous if the hardware effects were eliminated and function matching on break points was carried out, i.e. frequencies were formally translated into one frequency dispersion range.

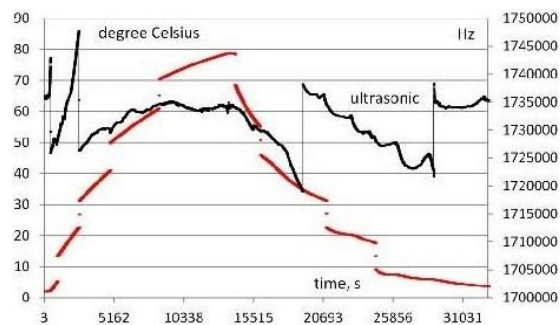

Fig. 4 A diagram of synchronous measurements of temperature (100 ml of distilled water heated by a 3W power heater) and resonance frequency of liquid in axial resonator

Results of this transformation are presented in Fig.5. Simultaneously frequency variations were translated into variations of ultrasound phase velocity relative to tabular values V (20° C) = 1482.7 m/s.

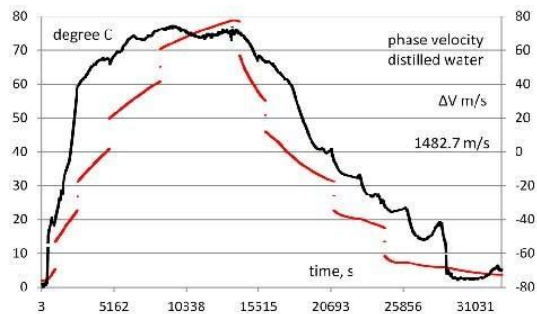

Fig. 5 Temperature-ultrasound velocity diagram, obtained synchronously in 100 ml of distilled water with a 3W power heater

Fig. 6 – 13 present fragments of the diagram (Fig. 5) in the vicinities of temperature indications jumps. Diagrams 5 – 13 have common time axis. Relation between the scales of temperature axis and acoustic measurements axis has been changed so that to make the comparison of the character of functional relationships more obvious.

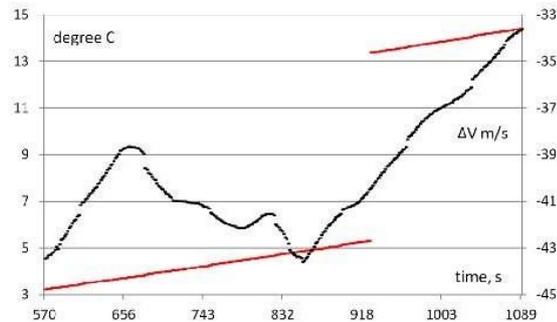

Fig. 6 Fragment of ultrasound velocity-temperature diagram in the area of temperature jump in case of synchronous measurements in total volume for the time interval of 570 – 1089 s

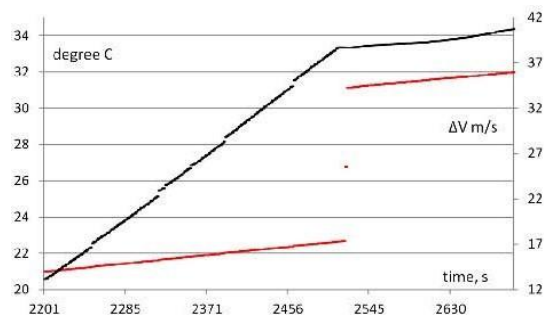

Fig. 7 Fragment of ultrasound velocity-temperature diagram in the area of temperature jump in case of synchronous measurements in total volume for the time interval of 2201 – 2700 s

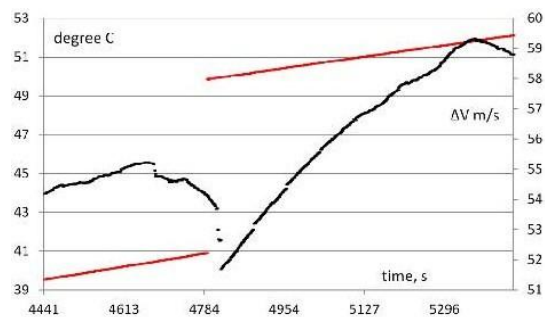

Fig. 8 Fragment of ultrasound velocity-temperature diagram in the area of temperature jump in case of synchronous measurements in total volume for the time interval of 4441 – 5450 s

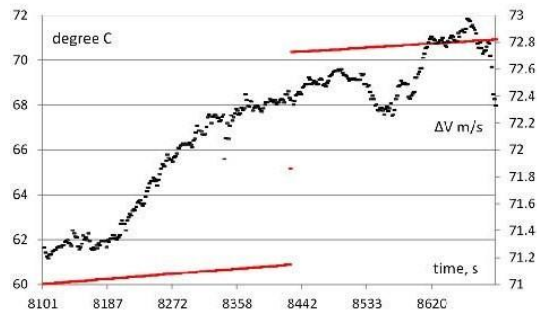

Fig. 9 Fragment of ultrasound velocity-temperature diagram in the area of temperature jump in case of synchronous measurements in total volume for the time interval of 8101 – 8700 s

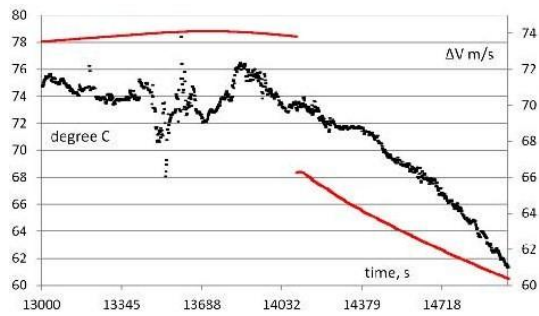

Fig. 10 Fragment of ultrasound velocity-temperature diagram in the area of temperature jump in case of synchronous measurements in total volume for the time interval of 13000 – 15000 s

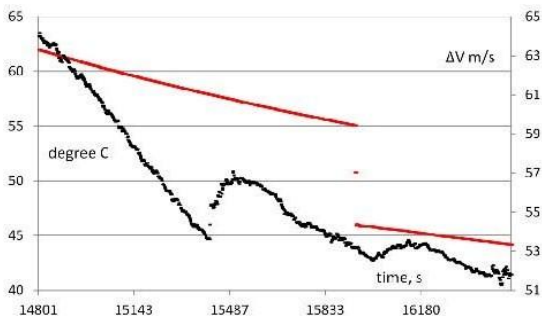

Fig. 11 Fragment of ultrasound velocity-temperature diagram in the area of temperature jump in case of synchronous measurements in total volume for the time interval of 14801 – 17000 s

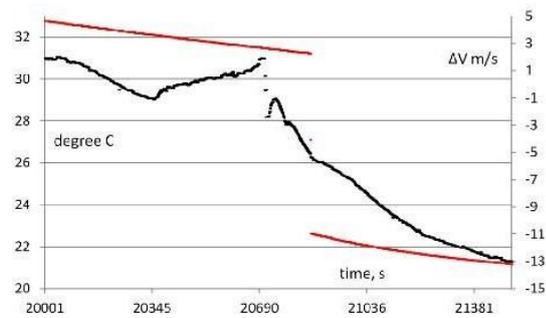

Fig. 12 Fragment of ultrasound velocity-temperature diagram in the area of temperature jump in case of synchronous measurements in total volume for the time interval of 20001 – 21400 s

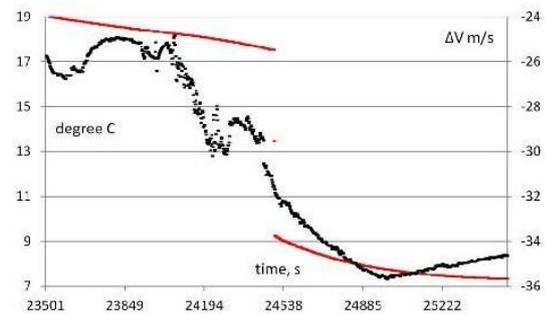

Fig. 13 Fragment of ultrasound velocity-temperature diagram in the area of temperature jump in case of synchronous measurements in total volume for the time interval of 23501 – 25700 s

Comparison of diagrams 6 – 13 makes it possible to state correspondence in time of stepped temperature transitions and substantial changes of resonance frequency of ultrasonic vibration. Reasons of these synchronous non-equilibrium phase transitions may be related to general nature of the observed processes. Experimental conditions of recording acoustic parameters of the medium are to a certain degree similar to the conditions of temperature tests: resonator – similar to resistive temperature detector – generates alternating electric field in the medium and also satisfies the requirement of galvanic insulation. A certain time shift of these processes is observed. In order to investigate the time shift, the experimental conditions were changed: an insulated copper wire, 0.05 mm in diameter, in the form of a bifilar winding was used as a temperature sensor; the wire length corresponded to the resistance of 30 Ohm. Temperature and acoustic measurements of distilled water in heating mode were reiterated in accordance with the above technique. Fig. 14 presents results of the corresponding synchronous measurements of temperature.

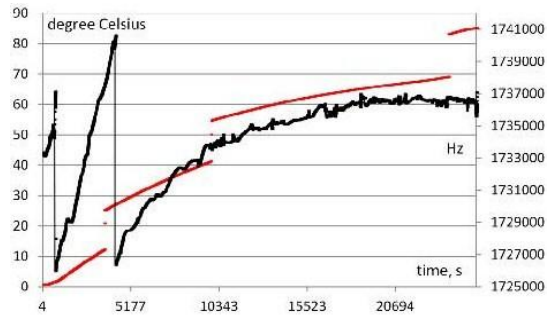

Fig. 14 Temperature-time and resonance frequency-time diagram

Nature of resonance frequency-time dependence has not considerably changed as compared to similar dependence, presented on Fig. 4. Features of frequency variations recurred on scale of temperature. Behavior of temperature indications of copper sensor – when medium was heated in accordance with the linear law – has changed in essence: synchronism in variations of acoustic parameters and temperature has not been observed. Fig. 15 presents a fragment of present dependencies in the vicinity of non-equilibrium phase transition. Variations of acoustic signal in the area of temperature jump are missing.

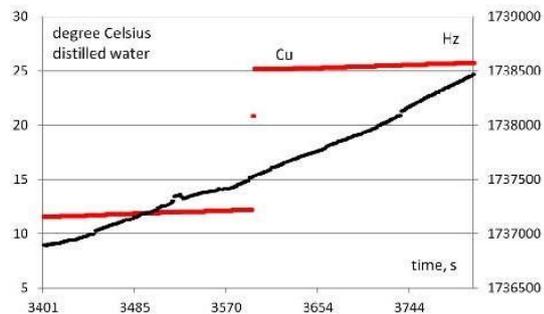

Fig. 15 Fragment of ultrasound frequency-temperature diagram (area of temperature jump, measured by a copper sensor for the time interval of 3401 – 3800 s)

It can be concluded that phase transition parameters are substantially determined by the geometry of the surface. Close agreement of variations of the results of acoustic measurements and temperature jumps in previous experiments only reflects the fact that flat surface of sensor HEL-775-B ($1mm^2$) and cylindrical surface (R = 8 mm) differ from each other in curvature to a very little degree.

Consistent explanation of the results of the experiments may be given within the framework of the model of surface-bound liquid layer and the evolution of dissipative processes, occurring in this layer in case of slow thermal energy inflow from the environment [1 – 3].

Paper [2], devoted to investigation of surface-bound water layer, gives a careful consideration of relationship between surface curvature and physical properties of the boundary layer: in case of convex surface liquid is more compressed, i.e. it experiences greater (positive) pressure, whereas in case of concave surface liquid is less compressed (negative pressure relative to the flat surface). This affects evolution of dissipative processes in surface-bound liquid layer.

Results of the experiments indicative of essential dependence of parameters of non-equilibrium phase transitions on the surface condition of the resistive structure were supplemented by an experiment with activated carbon. Activated carbon, used for medical purposes (0.25g, active surface area ~250m$^2$) and water were placed into a medical syringe and packed together with the electrodes. The amount of water in the carbon volume was defined by weighing and made 0.25g. The resulting resistive structure (~1300 O) was placed into a heated water volume (100ml, 3W). Sensor readings of temperature were registered in accordance with the technique described above (sensor resistance was lowered alongside with the temperature growth). Fig. 16 presents sensor readings in the process of heating and cooloing of water volume. Apparent regular alternations of temperature jumps can be observed. Parameters of phase transitions considerably differ from the results presented above. This is a complementary verification of the surface condition effect on the quantitative characteristics of such transitions. It should be noted that in the present experiment sensor dimensions considerably exceeded the dimensions of the resistive structures described above. Still, alongside with this, time intervals of temperature transitions remained rather small: 1 – 2s.

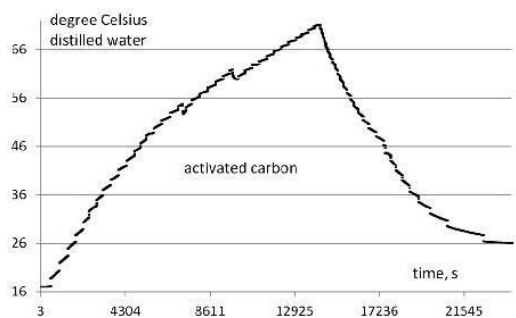

Fig. 16 Non-equilibrium phase transitions, registered by a resistive sensor in the activated carbon-water mixture

Aiming at further study of the role of surface in the process of forming non-equilibrium phase transitions, the following experiment was carried out. For this experiment a resistive sensor in the form of a ceramized substrate with a vacuum deposited metal layer was used. The sensor area amounted to 480 mm², resistance ~0.9O. Heating and measurements were carried out in accordance with the technique described above. Fig. 17 presents this temperature-time dependence.

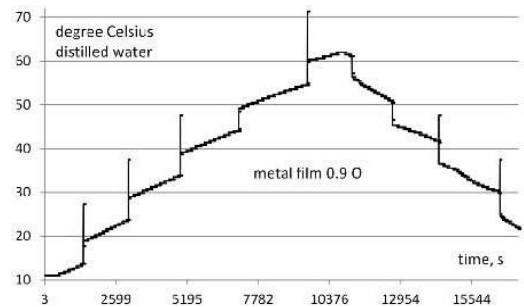

Fig. 17 Non-equilibrium phase transitions, registered by a resistive sensor, deposited on a ceramized substrate

The dependence illustrated by Fig. 17 can be interpreted in the following way. In the process of heating the surface-bound layer intensively absorbs energy from the environment which results in local temperature drop in the area of the resistive sensor. Owing to this, the registered temperature is different from the quasi-equilibrium value. On reaching the conditions of phase transition the molecular structure of two-dimensional crystal is transformed which results in release of accumulated energy into the environment. Due to high thermal conductivity this energy is absorbed at once by the metal film. A temperature drop of ~15K is registered. Then this heat dissipates in the substrate and surrounding liquid. After the process is completed the sensor temperature comes to equilibrium with the environment. Then the process is repeated.    In case of the medium getting cold we can observe a similar pattern on the diagram, i.e. structural dynamics of dissipative phenomena in the two-dimensional crystal.

Fig. 6 – 8 illustrate a peculiarity in variations of the phase velocity of the acoustic signal in the temperature range of 5 - 50°C. Repetitive discontinuity of a function can be observed. The value of the discontinuity of a function amounts to ~0.5m/s in velocity scale or $\Delta f \sim 10^{-4} f_{osz}$ in frequency scale where $f_{osz}(21.6^{O}C) = 1.761MHz$. It can be assumed that the observable fine structure of variations of ultrasound phase velocity is related to temperature by way of temperature changes of natural resonance frequency of liquid volume. Coherent ultrasonic field changes the condition of the near-surface layer of the medium. As a result interference of the wave acoustic field and surface-bound layer of the ordered medium can be observed. Similar results were obtained (Fig. 18) in the process of studying the influence of coherent ultrasonic field on liquid media by constant temperature [6]. Fig. 18 presents resonance natural frequency-time dependence (2% glycine

solution in distilled water by 20° C). A multiple-beam interferometer with plane piezoelectric transducers [4] ~ 2 ml in volume was used for carrying out the measurements. A similar picture of medium relaxation under the influence of coherent ultrasonic field could be also observed for other liquids [6].

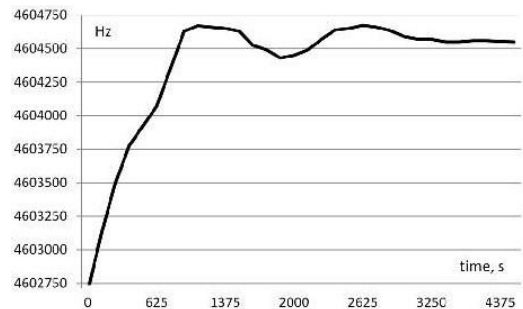

Fig. 18 Ultrasonic frequency-time dependence (under the influence of coherent ultrasonic field on 2% glycine solution by 20°C)

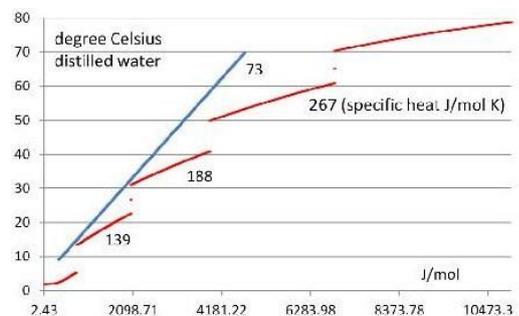

Fig. 19 Temperature-absorbed energy dependence

Fig. 19 presents a diagram of temperature (measured by sensor HEL-775-B)-absorbed energy (in the water volume) dependence. (It is a fragment of the diagram, presented in Fig. 4.) Heat capacity calculation data, presented on the diagram, correspond to the tabulated values only for the envelope functional relationship. The rest of the data should be interpreted subject to the non-equilibrium conditions of temperature registration.

Structural evolution of dissipative processes with external energy afflux is called "non-equilibrium phase transitions" by a number of authors [7, 8]. Non-equilibrium phase transitions were observed in all types of liquid media, studied in the present paper with the help of sensor HEL-775-B. Basic results of this research are presented in Tables 1 and 2.

Table 1 contains data for melt and chilled distilled water, sea salt solutions of different concentration and lymph substituting liquid. The table presents the initial temperature values, time interval of the complete cycle "heating – cooling", temperature drops intervals and their duration. Table 2 presents similar data for non-water liquid media. In all the experiments, except the ones with Hg, heating was carried out with sustained power. Cooling mode did not match the condition of linear dependence of heat energy extraction on time.

Table 1

| Distilled water, ice. Initial temperature 0.64°C. Measuring span, s 33962 | Distilled water, chilled. Initial temperature 0.38°C. Measuring span, s 28399 | Lymph substituting liquid Hemodez-H. Initial temperature -0.33°C. Measuring span, s 24670 | Sea salt solution 0,85 g/l. Initial temperature -3.1°C. Measuring span, s 21851 | Sea salt solution 3,45g/l. Initial temperature -12.36°C. Measuring span, s 27827 | Sea salt solution 330 g/l. Initial temperature -13.38°C. Measuring span, s 28522 |
|---|---|---|---|---|---|
| 1.48/9.11 (7.63 K, 1 s) | 6.24/ 14.12 (7.88 K, 1 s) | 2.46/10.41 (7.95 K, 2 s) | 2.07/9.83 (7.76 K, 2 s) | -4.61/2.84 (7.45 K, 4 s) | -4.68/2.87 (7.65 K, 4 s) |
| 17.91/26.15 (8.24 K, 1 s) | 22.95/31.24 (8.29 K, 5 s) | 18.47/26.79 (8.32 K, 4 s) | 15.07/23.13 (8.06 K, 1 s) | 2.18/9.95 (7.77 K, 3 s) | 10.98/18.94 (7.96 K, 3 s) |
| 35.77/44.42 (9.05 K, 3 s) | 41.2/50.05 (8.85 K, 5 s) | 36.62/45.37 (8.75 K, 1 s) | 33.44/41.97 (8.53 K, 2 s) | 16.53/24.59 (8.06 K, 2 s) | 28.10/36.49 (8.39 K, 2 s) |
| 56.29/65.53 (9.24 K, 3 s) | 64.76/74.00 (9.24 K, 1 s) | 39.71/31.06 (8.65 K, 3 s) | 49.51/40.74 (8.77 K, 4 s) | 34.34/42.86 (8.52 K, 3 s) | 47.54/56.35 (8.81 K, 2 s) |
| 64.92/55.73 (9.19 K, 3 s) | 64.92/55.66 (9.26 K, 3 s) | 24.69/16.45 (8.24 K, 3 s) | 36.00/27.56 (8.44 K, 2 s) | 49.56/40.74 (8.82 K, 3 s) | 53.94/44.98 (8.96 K, 2 s) |
| 42.66/33.93 (8.73 K, 3 s) | 46.52/37.77 (8.75 K, 3 s) | | 22.65/14.56 (8.09 K, 2 s) | 36.16/27.76 (8.40 KI, 3 s) | 23.95/15.86 (8.09 K, 2 s) |
| 25.67/16.97 (8.7 K, 1 s) | 31.91/23.49 (8.42 K, 3 s) | | | 22.06/13.97 (8.09 K, 3 s) | |
| | 11.95/4.07 (7.88 K, 3 s) | | | | |

Table 2

| Glycerine<br>Initial temperature<br>-9.7$^O$C.<br>Measuring span, s<br>12403 | Mineral oil<br>Initial temperature<br>-8.03$^O$C.<br>Measuring span, s<br>9501 | Carbon tetrachloride<br>Initial temperature<br>-2.76$^O$C.<br>Measuring span, s<br>8649 | Hydrargyrum<br>Initial temperature<br>-14.23$^O$C.<br>Measuring span, s<br>30259 |
|---|---|---|---|
| -7.42/0.05<br>(7.47 K, 1 s) | -4.45/3.12<br>(7.57 K, 2 s) | 1.23/9.19<br>(7.96 K, 3 s) | -4.79/2.97<br>(7.76 K, 2 s) |
| 9.16/17.04<br>(7.88 K, 1 s) | 11.77/19.73<br>(7.96 K, 3 s) | 16.79/25.10<br>(8.31 K, 4 s) | 15.51/24.33<br>(8.82 K, 3 s) |
| 26.41/34.75<br>(8.34 K, 4 s) | 28.86/37.31<br>(8.45 K, 1 s) | 33.34/42.14<br>(8.80 K, 6 s) | 40.38/49.69<br>(9.31 K, 7 s) |
| 44.34/53.12<br>(8.78 K, 1 s) | 47.36/56.29<br>(8.93 K, 4 s) | 51.69/42.30<br>(9.39 K, 3 s) | 58.01/67.78<br>(9.77 K, 4 s) |
| 46.29/37.69<br>(8.60 K, 4 s) | 55.17/45.47<br>(9.7 K, 3 s) | 33.39/24.49<br>(8.90 K, 3 s) | 71.06/80.83<br>(9.77 K, 3 s) |
| 31.91/23.67<br>(8.24 K, 3 s) | 36.54/28.15<br>(8.39 K, 4 s) | 18.35/9.47<br>(8.88 K, 4 s) | 93.60/104.15<br>(10.55 K, 4 s) |
| 21.65/14.51<br>(7.14 K, 7 s) | 23.11/14.97<br>(8.14 K, 3 s) | | 117.76/128.76<br>(11.00 K, 3 s) |
| | | | 136.77/148.75<br>(11.98 K, 4 s) |
| | | | 143.91/131.96<br>(11.95 K, 3 s) |
| | | | 120.70/109.39<br>(11.31 K, 3 s) |
| | | | 97.80/87.23<br>(10.57 K, 1 s) |
| | | | 75.79/66.04<br>(9.75 K, 1 s) |
| | | | 54.25/45.21<br>(9.04 K, 1 s) |
| | | | 28.35/20.01<br>(8.34 K, 3 s) |

It can be assumed that the reason of non-equilibrium phase transitions in non-water liquid media is the presence of a surface-bound water layer on the resistive sensor.

The experiments demonstrated that the temperature of non-equilibrium phase transitions depends on the prehistory of liquid medium condition. Fig. 20 presents the dynamics of temperature transitions in case of considerable variations of the rate of storage and consumption of the medium heat energy. Alongside with this, the amplitude of temperature drops does not change and remains within the margins of measurement errors. As follows from the data presented in Tables 1, 2, the amplitude of temperature drops grows concurrently with the growth of absolute temperature. Under invariant starting position, the dynamics of non-equilibrium phase transitions sequence is reproduced accurately within the margins of measurement error.

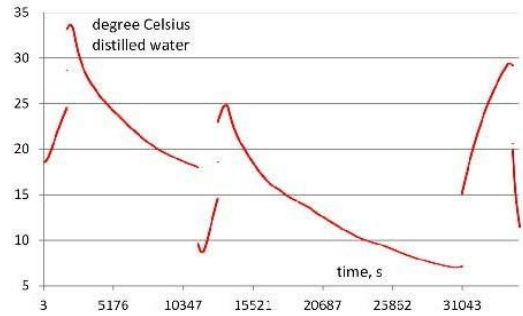

Fig. 20 Temperature-time dependence, illustrating the relationship between the parameters of non-equilibrium phase transitions and prehistory of the liquid medium condition

**Conclusion**

The proposed method [4, 6] of recording ultrasonic parameters of liquid media made it possible to discover non-equilibrium phase transitions in liquid structures in the process of changing the medium thermal energy. Simultaneously synchronous stepped variations of temperature were recorded by a film resistive sensor. It has been discovered that synchronism is disturbed in case of notable difference in surface curvature of acoustic and temperature sensors. The observed effects exist in case of imposing specified conditions in relation to the equipment used for carrying out the research. The stepped temperature dependence generated by the dissipative medium is created in presence of alternating electric field under conditions of slowly varying control parameter – temperature. The results of the present investigation reflect, generally, the behavior of an open macroscopic system. It is known that in such systems there can be generated different types of structures, in the formation of which dissipative processes play a constructive part. Making a physical analog explaining the experimental results of the present paper is the task of further research. Possible application of the attained results may be usage of the proposed methods for making experiments on studying a two-dimensional molecular structure of a water layer, adjacent to solid surface. It can be assumed that the most interesting application of the proposed technique may be cellular structures of living systems. Existence of these systems is based on space-time non-equilibrium processes with dissipative losses, compensated by energy gain [9].